\documentclass{article}
\usepackage[utf8]{inputenc}

\usepackage{hyperref}
\usepackage{graphicx}
\usepackage{tikz}
\usepgflibrary{decorations.shapes}
\usepackage{pgfplots}
\usepackage{float}
\usepackage{xcolor}
\pgfplotsset{compat=1.18}

\usepackage{caption} 
\usetikzlibrary{arrows}

\definecolor{rosso}{RGB}{229,124,0}
\definecolor{giallo}{RGB}{198,211,37}
\definecolor{blu}{RGB}{0,177,234}
\definecolor{verde}{RGB}{0,108,102}
\definecolor{viola}{RGB}{119,119,119}
\definecolor{red}{RGB}{53,20,103}
\definecolor{adjusted_red}{RGB}{108, 25, 0}

\tikzset{
  chart/.style={
    legend label/.style={font={\scriptsize},anchor=west,align=left},
    legend box/.style={rectangle, draw, minimum size=5pt},
    axis/.style={black,semithick,->},
    axis label/.style={anchor=east,font={\tiny}},
  },
  pie chart/.style={
    chart,
    slice/.style={line cap=round, line join=round, very thick,draw=white},
    pie title/.style={font={\bfseries}},
    slice type/.style 2 args={
        ##1/.style={fill=##2},
        values of ##1/.style={}
    }
  }
}

\pgfdeclarelayer{background}
\pgfdeclarelayer{foreground}
\pgfsetlayers{background,main,foreground}

\newcommand{\pie}[3][]{
    \begin{scope}[#1]
    \pgfmathsetmacro{\curA}{110}
    \pgfmathsetmacro{\radius}{1}
    \def\Centre{(0,0)}
    \node[pie title] at (80:1.4) {#2};
    \foreach \v/\s in{#3}{
        \pgfmathsetmacro{\deltaA}{\v/100*360}
        \pgfmathsetmacro{\nextA}{(\curA) + (\deltaA)}
        \pgfmathsetmacro{\midA}{(\curA+\nextA)/2}

        \path[slice,\s] \Centre
            -- +(\curA:\radius)
            arc (\curA:\nextA:\radius)
            -- cycle;  

  \pgfmathsetmacro{\ysign}{ifthenelse(mod(\midA,360)<=180,1,-1)}
  \pgfmathsetmacro{\xsign}{ifthenelse(mod(\midA-90,360)<=180,-1,1)}

   \begin{pgfonlayer}{foreground}
       \draw[-,thin] \Centre ++(\midA:\radius/1.275) -- 
                             ++(\xsign*0.07*\radius,\ysign*0.2*\radius) --
                             ++(\xsign*0.5*\radius,0) 
                      node[above,near end,pie values,values of \s]{$\v\%$};
   \end{pgfonlayer}

        \global\let\curA\nextA
    }
    \end{scope}
}

\newcommand{\legend}[2][]{
    \begin{scope}[#1]
    \path
        \foreach \n/\s in {#2}
            {
                  ++(0,-10pt) node[\s,legend box] {} +(5pt,0) node[legend label, text width=4cm] {\n}
            }
    ;
    \end{scope}
}

\title{Handling Open Research Data within the Max Planck Society -- Looking Closer at the Year 2020}
\raggedright
\author{
    Martin Boosen \url{https://orcid.org/0009-0009-3989-5087},\\
    Michael Franke
    \url{https://orcid.org/0000-0002-2661-8242},\\
    Yves Vincent Grossmann \url{https://orcid.org/0000-0002-2880-8947},\\
    Sy Dat Ho \url{https://orcid.org/0000-0002-6218-4146},\\
    Larissa Leiminger \url{https://orcid.org/0000-0002-6491-3197}, and \\Jan Matthiesen \url{https://orcid.org/0000-0001-6548-3654}\\
    Max Planck Digital Library \url{https://ror.org/0061msm67}\\
    \href{mailto:rdm@mpdl.mpg.de}{rdm@mpdl.mpg.de}
}

\date{February 2024}

\usepackage[style=authoryear,]{biblatex}
\begin{document}
\begin{titlepage}
\maketitle

\begin{abstract}
This paper analyses the practice of publishing research data within the Max Planck Society in the year 2020. The central finding of the study is that up to 40\% of the empirical text publications had research data available. The aggregation of the available data is predominantly analysed. There are differences between the sections of the Max Planck Society but they are not as great as one might expect. In the case of the journals, it is also apparent that a data policy can increase the availability of data related to textual publications. Finally, we found that the statement on data availability "upon (reasonable) request" does not work.
\end{abstract}

\end{titlepage}

\pagebreak

\tableofcontents

\pagebreak

\section{Introduction}
Data sharing is becoming a new standard in research. Data accessibility is increasingly being discussed as an important aspect of scientific reproducibility. Funding agencies and public authorities are increasingly demanding that their money be used to make research data as openly accessible as possible. Many efforts are being made to ensure that the conscious handling of research data becomes an established practice in everyday scientific life.\\
However, there is sometimes a gap between aspiration and reality. The reasons for this are to be found on the side of the scientists, the publishers, and on the side of the infrastructure providers. Although there is often some kind of agreement -- voluntary or involuntary -- to share data. But it is still remarkably rare. To change this, the aim would be to take the pain out of data sharing.\footnote{See for example the recently published overview \cite{hutson_taking_2022}.} This and other debates around data sharing are discussed with reference to the Max Planck Society in the following chapters.\newline

As already mentioned data sharing is perceived as an important element of intra-scientific exchange. At the same time, the understanding of data sharing is hampered by different interpretations of the pair "data" and "sharing".
Focusing on research data from the perspective of the Max Planck Society as well as on its practices this paper acknowledges two ways of understanding this term. On the one hand, research data is the basis for scientific findings and their publication.\footnote{See the old "Rules of Good Scientific Practice" of the Max Planck Society from 2000 resp. 2009, \cite[p. 4]{max-planck-gesellschaft_regeln_2009}.} On the other hand, published research data is increasingly emerging as a genre of scholarly output in its own right. The two come together again when we ask about the accessibility of the data underlying published texts.\newline

The two sides of the research data coin -- wanting to commit to data sharing, but failing to do so -- are the focus of the first section. A general spectrum of data sharing is briefly outlined. The main emphasis is on developments since 2010. Here, the positions and initiatives involved in bringing data accessibility more into focus are traced. The second section examines the perspective of the Max Planck Society by presenting the open data and sharing practices of the decade of the 2010s. The main aim is to show which institutional frameworks have been established and which intentional vagueness (in the sense of a desired added value) has been maintained. The third section takes a closer look at the Max Planck Society and shows what its publication behaviour looks like in 2020 in terms of research data. This evaluation will take up the most space, as it formulates the data collection and the descriptive analysis of the results. The paper concludes with an outlook on the decade of the 2020s and possible developments in data sharing for the Max Planck Society.\pagebreak

\section{Observations on Sharing Research Data}

Data sharing is not a phenomenon of the 21$^{st}$ century. Sharing knowledge has long been an ingrained cultural practice of mankind. At the same time, data sharing seems to be on the rise since the beginning of the third millennium, not least because of the possibilities digitalization brings. This development is new because it has given rise to a new genre of publication called "data" or "dataset". This data sharing is accompanied by several phenomena.\newline

First and foremost, the open exchange of data is becoming increasingly important in this context. Open means barrier-free access to the data. This is usually done by attaching the data to the textual publication or by publishing the data as a separate dataset in a repository. Especially in the latter case, there are already signs that data publishing is becoming a genre in its own right. This goes hand in hand with increased visibility of research data. They can gain visibility through open or restricted access options. Nevertheless, they are accessible, or at least there is knowledge of their existence. Therefore, especially for research data with restricted access, it is important that the metadata, i.e. the descriptive information about the data, is freely available. This makes it much easier and faster for other researchers to learn of the existence of the research data and then request access. This is adding to the recognition as an entity in their own right.\newline

Regardless of this, data can also be reused in ways that were not foreseen by the original authors. The more conscious use of data licences facilitates such reuse and reinterpretation. Particularly in a scientific context, future questions cannot always be guessed at from current perspectives. That is why it makes sense to present research data in such a way that new questions can be discussed in the future. Parallel to software, licensing is increasingly becoming a familiar process. This creates legal certainty, which makes access easier -- in a legal sense. In many cases, this increase in visibility is accompanied by an increase in citations by other scientific results. It is therefore not surprising that the bibliometric study by Colavizza and colleagues found that there is "\textit{a citation advantage, of up to 25.36\% (± 1.07\%), with articles that have a} [..] \textit{link to a repository via a URL or other permanent identifier}".\footnote{\cite[p. 14]{colavizza_citation_2020}.}\newline

For the phenomena of sharing data openly, we see -- in view of Germany -- a strong acceleration of discussion contributions in the 2010s, but an increasing acceptance of the forms of data publication. For this reason we present a chronology for the recent past. An important milestone and turning point are the FAIR Data Principles\footnote{\cite{wilkinson_fair_2016}.}, which were published in 2016. Not only do they cover the decade of the 2010s, but they also provide explicit guidance on the publication of research data. At the same time, for the first time, the FAIR Principles provide a simple and concise way of articulating -- FAIR -- what has often been expressed in more descriptive terms when dealing with research data.\footnote{Due to the abundance of introductions, handouts and presentations, an explanation of the FAIR principles will not be given here. For the German context, however, the following is a suitable overview see \cite{go_fair_2021}.} Such a German frame of reference is naturally integrated into European and worldwide developments. Research is mostly international. Nevertheless, in the area of normative discussions, the Max Planck Society mainly operates in Germany, so this is the primary focus of this text.\newline

To reflect these developments, the following chapter discusses the evolution prior to the FAIR Data Principles 2016. This is followed in chapter \ref{Before-2016} by a description of the FAIR acceleration phase from 2016 to the present. These two parts serve as a derivation to the concrete study year 2020.
We will then focus in chapter \ref{After-2016} on specific aspects. They become particularly relevant again in the following chapter, i.e. the empirical analysis of publication behaviour. First, the text focuses on the particular aspect of data availability statements to illustrate how the approach to data accessibility has changed. However, this would be difficult to explain without the institutional framework. Therefore, the following two subsections discuss developments in institutional data policies in \ref{Research-Data-Policies} and funders \ref{Third-Party} to conclude the general perspective on research data sharing. Finally, this chapter also serves as a general introduction and transition in terms of content and timing for the next parts of the text.

\subsection{Discussion on Sharing Research Data before 2016} \label{Before-2016}

Looking back, it is often difficult to pinpoint the exact start of a discussion. Therefore, temporal categorisations often have something artificial or arbitrary about them. The same is true here. Nevertheless, the dilemma remains that start and end points must be specified. One of the first publications with specific recommendations worldwide on research data was a 53-page publication by the OECD in 2007.\footnote{\cite{oecd_oecd_2007}. Particularly worth reading are Principles A to M, which attempt to define concrete guidelines for the design of data access.} In particular, the aim was to develop a set of guidelines, based on commonly agreed principles, to facilitate cost-effective access to publicly funded digital research data.\footnote{\cite[p. 3]{oecd_oecd_2007}.} The sequencing of human DNA -- successfully completed in 2003 -- was used to study many biological processes, so open sharing of research data was already implicit in the project.\newline

In the German context, the first significant contribution to the discourse can be seen in 2009. One of the DFG's first decentralised recommendations for action on research data was the "Empfehlungen zur gesicherten Aufbewahrung und Bereitstellung digitaler Forschungsprimardaten" of January 2009.\footnote{\cite{ausschuss_fur_wissenschaftlicheempfehlungen_2009}. An English translation of this would be "Recommendations for the Secure Preservation and Provision of Digital Research Primary Data".} This was shortly followed by the Alliance of German Science Organisations' "Principles for the Handling of Research Data" in 2010.\footnote{\cite{allianz_der_deutschen_wissenschaftsorganisationen_principles_2010}.} Both recommendations were intended to set initial standards for the concrete handling of research data.\newline

The "Kommission zur Zukunft der Informationsinfrastruktur"\footnote{A translation of this may be "Commission on the Future of the Information Infrastructure.}, the first joint institution with the focus on research data in the Federal Republic of Germany was established in 2009 on the recommendation of the Gemeinsame Wissenschaftskonferenz (GWK). This commission, with the participation of many German scientific organisations and institutions, also had its own working group on research data.\footnote{\cite[line 490-506]{kommission_zukunft_der_informationsinfrastruktur_gesamtkonzept_2011}.} This was followed in 2012 by recommendations from the German Science and Humanities Council (WR), which called for the professional communities or actors in interdisciplinary research fields to develop quality criteria for the generation of research data and guidelines for appropriate data management, where these do not already exist.\footnote{\cite[p. 56]{wissenschaftsrat_empfehlungen_2012}.} As an interim result, it was noted in 2015 that the field of research data is a very dynamic and diverse field.\footnote{\cite[p. 2]{franke_positionspapier_2015}.}\newline

In addition to these recommendations for action by institutional stakeholders, research and education institutions have been issuing their own data policies since the beginning of the 2010s. Based on the recommendations mentioned above, it can be observed for the German context that the number of data policies and universities and research institutions has increased significantly since 2014.\footnote{The best overview of research data policies in the German-speaking world is provided by \url{www.forschungsdaten.org/index.php/data\_policies}. It is worth mentioning the recommendations of the German Rectors' Conference (Hochschulrektorenkonferenz) of 2014, \cite{hochschulrektorenkonferenz_management_2014}. The two-year period between the WR recommendation and the first policies can be reasonably explained by an internal development and establishment process. The introduction of a data policy in an institution simply requires different resources and time.} Already in 2011, the German debate on data policy is characterised by opposite poles of recommendation -- as recommendations in the hope that they will be followed out of conviction -- and obligation -- so real commitments.\footnote{For the German context, see in particular \cite{pampel_data_2011}.} This also fits in well with the fact that in Horizon 2020 (running from 2014 to 2020) the European Commission has for the first time called for a conscious handling of data, for example through data management plans (with the possibility to opt-out).\footnote{\cite{european_research_council_guidelines_2017}.} In retrospect, the long-term development towards obligatory requirements regarding research data at Horizon Europe is already indicated here.\newline

With these developments towards active management of research data, the positive aspects of data sharing and citation are being observed. Publishers have also begun to adopt data policies and request research data associated with submitted articles.\footnote{An example would be PLOS Biology, with the intention of increasing data availability and transparency, \cite{bloom_data_2014}.} In this context, we are beginning to see how a change in publishing behaviour -- towards data sharing -- is affecting the behaviour of scientists. For example, for some supporters, the willingness to share research data is linked to the strength of the evidence and the quality of reporting of statistical results.\footnote{\cite{wicherts_willingness_2011}.}\newline

Some evaluations have shown that the behaviour of scientists is moving towards open sharing of research data. "\textit{However, there is increased perceived risk associated with data sharing, and specific barriers to data sharing persist.}"\footnote{\cite[p. 1]{tenopir_changes_2015}.} It is interesting to note that the scientific communities have given themselves their own guidelines or recommendations on how best to deal with research data in their environment.\footnote{See for example \cite{goodman_ten_2014}.} For some aspects, this development can be repeated as well at a general level, towards universal principles for the handling of (research) data.

\subsection{Understanding Data Sharing Practices after Announcing the FAIR Data Principles in 2016} \label{After-2016}

The publication of the FAIR Data Principles\footnote{\cite{wilkinson_fair_2016}.} in 2016 can, in retrospect, be seen as an important point on the way to an evolving awareness of data sharing. Two things in particular were central to this. First, there was broad acceptance of the principles. And secondly, it was now possible to give a name to both the process and the outcome, namely FAIR. This naming, as an explicit accumulation of knowledge, should not be underestimated in the success of the FAIR Principles.\footnote{A general overview of the introduction of the FAIR Data Principles in the first years can be found in \cite{thompson_making_2019}. At the same time, this already contains an existing criticism of the FAIR Data Principles, namely that they are often mentioned performatively, but the concrete application sometimes leaves much to be desired.}\newline

The FAIR Data Principles were adopted quite rapidly by the European Commission in 2016 and documented in the "Guidelines on FAIR Data Management in Horizon 2020"; \footnote{\cite{european_commission_h2020_2016}. See also later the report on turning FAIR into reality from 2018, \cite{european_commission_turning_2018}.} However, it did not stop at the introduction of guidelines. Already in 2017, the Open Research Data Pilot covered all thematic areas of Horizon 2020, which "\textit{aims to improve and maximise access to and re-use of research data generated by Horizon 2020 projects}".\footnote{\cite{european_commission_open_2018}.} The aim was to create incentives to convince scientists to organise their data according to the FAIR Data Principles.

Two years later, the European Commission produced an analysis of the cost of FAIR research data and the lack of application of the FAIR data principles. According to this document, "\textit{at €10.2bn per year in Europe, the measurable cost of not having FAIR research data makes an overwhelming case in favour of the implementation of the FAIR principles.}"\footnote{\cite{doi/10.2777/02999}.} From this perspective, the opportunity cost of not having the FAIR Data Principles was high. Beyond the EU as a major initiator, there have been many other developments towards the concrete application of the FAIR Data Principles.\newline

According to a survey conducted by the European Commission in 2022, recognition of research data management and sharing is already partly present, but could be significantly higher, according to the participants.\footnote{See in particular \cite[pp. 552-553]{neuroth_aktuelle_2021} with a geographical focus on Germany and similar discussions is particularly interesting in this context.} There are different approaches to encourage different actors to structure their data according to the FAIR Data Principles and, where possible, to make them publicly accessible. There are explicit commitments, e.g. through research data policies of institutions, deliverables from funders or demanded data availability statements by journal publisher. At the same time there are incentives to promote the conscious use of data, e.g. a data index\footnote{\cite{hood_data-index_2021}.} as a bibliometric measure, to help the independent genre of 'data publication' gain more recognition.\footnote{The development of data journals are an example of this.} There are many initiatives, ideas and new ways of doing things in the scientific communities that try to facilitate and promote the application of the FAIR Data Principles. However, it is difficult to generalise such developments in a meaningful way by subject or other criteria. Because "\textit{data sharing perceptions and practices are highly variable among academic disciplines.}" \footnote{\cite[p. 238]{pujol_priego_puzzle_2022}. In this whole discussion about the implementation of the FAIR Data Principles, however, it is striking that there is comparatively little explicit criticism of them.}\newline

Nevertheless, the genre of analysis of research data sharing behaviour has become common.\footnote{See also, for example, the meta-analysis of various studies \cite{donner_research_2022}.} There are two broad categories into which this can be grouped.
The first category is working on data sharing behaviour within a research field. Such publications are particularly concerned with understanding how data is handled within a discipline.\footnote{See for example \cite{borghi_data_2021}, \cite{cruwell_investigating_2022}, \cite{houtkoop_data_2018}, \cite{jeng_surveying_2022}, \cite{leonelli_global_2017}, \cite{mandeville_open_2021}, \cite{Rousi_2022} and more.} For example, to what extent is open sharing of research data a quasi-standard there? Or are there good reasons for restricting access to data? Are the FAIR data principles widely applied? How is the reproducibility of research data valued within the scientific community? And where is research data published? The overall aim of these analyses is to understand what patterns of behaviour currently exist and evolve. These questions are often compared to a desired state. At the same time, there are also comparisons between different disciplines, as well as cross-disciplinary analyses.\footnote{See for example \cite{enwald_data_2022}, \cite{feger_yes_2020}, \cite{gabelica_many_2022}, \cite{tedersoo_data_2021}, \cite{thoegersen_researcher_2022} and more. This list could be extended. However, it should be clear that such analyses are being carried out and are available.}

The second category of analyses is primarily interested in data publication from an institutional perspective. This perspective is mostly taken by infrastructure providers such as libraries and data centres. It focuses on the question of where data should be published by researchers affiliated to their own institution.\footnote{Examples of such studies are \cite{borghi_identifying_2021}, \cite{quigley_role_2022}, \cite{read_data-sharing_2021}, \cite{van_gend_open_2022} and more.} In addition to the collection of bibliometric statistics, this is a motivation for the continuous improvement of institutional research data services. In the German context, the Charité Dashboard on Responsible Research\footnote {\url{https://quest-dashboard.charite.de}. See also \cite{iarkaeva_semi-automated_2022}.} is particularly worthy of mention. There, the institution's open data and code publications can be viewed in quasi-live mode. Most of these studies of data sharing behaviour focus on the issue of restricting access without justification. It is interesting to emphasise here that in many cases after 2016, the main focus was on the concrete implementation of the FAIR Data Principles.

\subsection{Data Availability Statements: Make Data "Accessible" on Reasonable Request} \label{available-statements-text}
"\textit{All that glisters is not gold.}"\footnote{William Shakespeare (1564–1616), Merchant of Venice, act II, scene 7.} The same sometimes applies for research data. In theory, reproducibility is part of good research. But in reality, reproducibility of data-based findings is sometimes difficult or almost impossible.
The stumbling block is often a data availability statement, which publishers now routinely require for scientific publications. The aim is to document where the data, on which the hypotheses of the publication are based, is available. The proportion of data availability statements varies. In some cases, the coverage of articles is quite high, depending on the discipline and on the journals' data policies. For example, in 2018, 93.7\% of 21,793 PLOS articles and 88.2\% of 31,956 BMC articles had data availability statements.\footnote{\cite[p. 7]{colavizza_citation_2020}.}\newline

Unfortunately, the existence of an availability statement says nothing about its content. Often, it is stated the data is "available upon reasonable request". In one reasoned study with N=1792 statements about "93\% authors either did not respond or declined to share their data" after a request.\footnote{\cite{gabelica_many_2022}.} As a comparison of preprints with their published versions has shown, that "\textit{data availability statements} [...] \textit{are a good first step, but are insufficient to ensure data availability}".\footnote{\cite[p. 2]{mcguinness_descriptive_2021}.} To be clear, there is a lot of frustration by getting research data with these statements of data availability that end up coming to nothing.\newline

As already indicated, data sharing practices and the availability of data on request vary between scientific disciplines.\footnote{See a meta-study such as \cite{tedersoo_data_2021}.} On the one hand, it is quite a challenge to work out the different disciplinary cultures in order to compare these different frames of reference. On the other hand, it is also unsatisfactory to postulate that everything is complicated without showing a higher level of detail. Ultimately, however, there are many different reasons why data sharing does not take place.\footnote{See generally \cite{gomes_why_2022}.} With a view to the Max Planck Society and the year 2020, we will take a closer look at some of these reasons in chapter \ref{Available-upon-request}.

\subsection{Open Research Data Policies in 2020} \label{Research-Data-Policies}

Open sharing of research data is not the norm and, unlike the application of the FAIR principles, it is not always appropriate in all areas. At the same time, many European scientific institutions have policies that provide legal certainty for the free sharing of research data.\footnote{But it should be clear that this is not a purely European phenomenon. Similar issues are being debated in other regions and countries. See for example the review of open research data policies in China \cite{zhang_review_2021}.} This issue is also increasingly being taken up by publishers, so that more and more data policies are being established.\footnote{See for example the overview \cite{hrynaszkiewicz_developing_2020}. For a general overview of research data policies in journals in the 2010s and their evolution, see in particular \cite{dearborn_changing_2018}.} Research organisations, such as the German Helmholtz Association and the French Institut Pasteur, have issued guidelines on the management of research data.\footnote{\cite{helmholtz_open_science_empfehlungen_2019} and \cite{pasteur_politique_2021}.} What is generally striking about these processes towards a data policy is that they require a lot of time and energy from all stakeholders. Internal coordination and the development of a common understanding are resource-intensive. Funders are well aware of this. The Swiss National Science Foundation, for example, evaluated the introduction of its Open Data Policy in 2020 and calculated the associated funding expenditure.\footnote{\cite[p. 4-5]{milzow_open_2020}.}\newline

For the Max Planck Society, it should be noted that there is no general research data policy for 2020. Individual institutes and departments have dealt with this issue in different ways. Based on the Harnack Principle, however, quite different solutions have emerged locally in the institutes and departments.\footnote{\cite{laitko_harnack-prinzip_2015}.} One example is the internal guideline on the handling of research data at the Max Planck Institute for the Study of Collective Goods. It was made mandatory for all employees in 2018.
\footnote{\cite{mpicoll_fdm-policy_2018}.} It defines research data, regulates ownership, documents local data management and agrees on the implementation of the regulations. Another example of a local policy is the Biomaterials Department at the Max Planck Institute of Colloids and Interfaces.\footnote{\cite{simon_datenmanagement_2020}.} This defines the procedures for handling samples in a departmental data policy.\\

At the same time, it should not be hidden that in many cases there are no explicit procedures, recommendations, etc. at the Max Planck Institutes. This can be seen as a sign that the existing workflows with research data are already carried out at a high level. At the same time, it should be noted that there is often an implicit procedure for research data. This does not always lead to a common approach to research data management and sometimes causes local problems.

\subsection{Perspective of Third-Party Funders on Open Research Data} \label{Third-Party}

In addition to this intrinsic motivation, there is also the perspective of funders on data sharing. There is a clear tendency to make open access to research data a condition of grant approval.
This applies not only to European, but to national funders. For the Max Planck Society, both the European Commission and the German Research Foundation were particularly important in 2020, each accounting for nearly a third of approved external funding.\footnote{\cite[p. 45]{max-planck-gesellschaft_jahresbericht_2021}} 

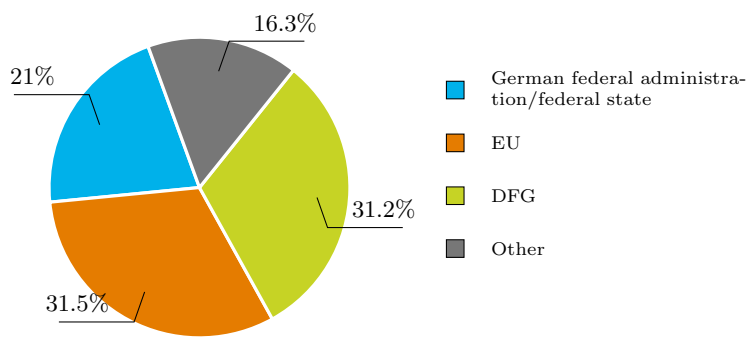
\begin{figure}[h!]
\centering
\begin{tikzpicture}
[
    pie chart,
    slice type={comet}{blu},
    slice type={legno}{rosso},
    slice type={coltello}{giallo},
    slice type={sedia}{viola},
    pie values/.style={font={\small}},
    scale=2.0
]

\pie{}{21/comet,31.5/legno,31.2/coltello,16.3/sedia}

\legend[shift={(1.7cm,1cm)}]{{German federal administration/federal state}/comet,{EU}/legno, {DFG}/coltello,{Other}/sedia}
\end{tikzpicture}
\caption{Distribution of third-party funds of the Max Planck Society in 2020}
\label{figure:mpg-jahresbericht}
\end{figure}

Research data has become increasingly important to the European Commission since the 2010s. Horizon 2020 explicitly includes research data and, in particular, data management plans.\footnote{For a good overview of these developments with static evaluations, see especially \cite[p. 49-63]{athenaresearchmonitoring2021}.} At that time, a data management plan (DMP) was not yet mandatory. At the same time, the availability of data management plans led to increased writing of such. In a highly competitive environment, it was no longer always possible to opt-out this element in relation to competitors. The already mentioned Open Research Data Pilot provided an opportunity for the Commission to promote the openness of research data.\footnote{See for example \cite{european_research_council_guidelines_2017}.} In addition, the Commission has repeatedly tried to point to new developments. Only two years after the publication of the FAIR Data Principles, an attempt was made to put them into practice by releasing concrete financial resources, not just statements of intent.\footnote{\cite{european_commission_turning_2018}.}\\

With the European Open Science Cloud (EOSC), the Commission tried early on to establish both a funded infrastructure for research data.\footnote{For a good and concise summary of the history of the EOSC, see in particular \cite[p.13]{rfii_foderierte_2023}. For a perspective on the EOSC for Max Planck scientists see also \cite{grossmann_european_2021}.} Open research data played a special role. On the part of the Max Planck Society, the MPCDF was particularly involved, especially at \href{https://eoscpilot.eu}{EOSCpilot} with its own science demonstrators.\newline

Compared to the European Commission, the DFG's developments in open research data up to 2020 are rather modest. This is primarily due to the self-governing nature of German science. Structurally, this is preceded by discussions within the discipline until, for example, a recommendation from a DFG commission is adopted. An early example of this is the 2015 Guidelines in the context of biodiversity research.\footnote{\cite{forschungsgemeinschaft_richtlinien_2015}.} In retrospect, the emergence of the German National Research Data Infrastructure (NFDI) at the national level is probably the point at which research data management gained importance in Germany, and in the DFG in particular. The first call for NFDI consortia was launched in 2019. In 2020, the first consortia were named and the second call was published.\footnote{For statistical evaluations of these two rounds, see in particular \cite{forschungsgemeinschaft_nationale_2020} and \cite{forschungsgemeinschaft_nationale_2020-1}.} For discussions within the scientific community, the developments around the NFDI on Open Research Data can hardly be overestimated.\footnote{For an evaluation of the first two NFDI calls with a focus on the participation of the Max Planck Society, see \cite{grossmann_participation_2021}.}\pagebreak

\section{Open Research Data in the Max Planck Society}
Open research data has been of declared importance within the Max Planck Society since the beginning of the millennium. The Berlin Declaration of 2003 also refers specifically to data in its definition of Open Access contributions.\footnote{"\textit{Open access contributions include original scientific research results, raw data and metadata, source materials, digital representations of pictorial and graphical materials and scholarly multimedia material.}" \cite{max-planck-gesellschaft_berliner_2003}.} In 2020, however, the Max Planck Society did not have a general data policy or other recommendations for dealing with research data. At the same time, this does not mean that research data, and in particular open access to them, are not important to Max Planck scientists. An analysis by the Max Planck PhDnet Open Science Group, for example, highlights the potential of open research data for early career researchers.\footnote{\cite[p. 3]{toribio-florez_where_2021}.} Such discussions within the Max Planck Society are always embedded in the normative framework provided by the rules of good scientific practice.

\subsection{Good Scientific Practices within the Max Planck Society}
In 2000, the Senate of the Max Planck Society adopted new rules to ensure good scientific practice.\footnote{\cite{max-planck-gesellschaft_regeln_2009}.} These were amended in 2009 and replaced by new rules in 2022 without widespread communication.\footnote{\cite{max-planck-gesellschaft_verhaltensregeln_2022}.} It is therefore important to bear in mind that the 2000 and 2009 regulations still applied in the year 2020. But it was already clear in 2020 that the regulations would change. This was initiated by the DFG and the adoption of the Guidelines for Safeguarding Good Research Practice, which became binding for applicant institutions such as the Max Planck Society.\footnote{\cite{deutsche_forschungsgemeinschaft_code_2019}.} The situation in 2020 was therefore characterised by an impending change in the normative framework for research data within the Max Planck Society.\footnote{For an Max Planck internal perspective on this situation see especially \cite{franke_forschungsdaten-policies_2020}.}\newline

In the year 2020 Max Planck Rules for Safeguarding Good Scientific Practice, research data (in German "Forschungsdaten") were not yet referred to by this term. Rather, it was usually referred to as data or primary data. Nevertheless, the data had to be processed according to discipline-specific rules. Primary data had to be kept for ten years. It was also necessary to ensure that clear and comprehensible documentation was given, for example in laboratory notebooks. Access to the data had to be guaranteed for authorised interested parties.\footnote{\cite[p. 2 and p. 4]{max-planck-gesellschaft_regeln_2009}.} However, there was already open research data. Various services for open research data are already existing in 2020.

\subsection{Max Planck Services for Open Research Data}
The central infrastructure and service units of the Max Planck Society have been offering open research data services for some time. Some services for open research data were therefore already available within the Society in 2020. Three services are briefly presented here as examples to document an impression at this point in time.\newline

In summer 2020, the MPCDF launched a CKAN instance for \href{https://pandoradata.earth}{Pandora}. This is a data repository for open archaeological data for the Max Planck Institute for the Science of Human History\footnote{The former Max Planck Institute for the Science of Human History was renamed the Max Planck Institute for Geoanthropology in 2022.} and its collaborators. The repository has since supported the Pandora initiative to make historical and archaeological data more accessible and discoverable.\newline

Since 2014, the Max Planck Digital Library offers \href{https://edmond.mpdl.mpg.de}{Edmond}, a repository for open research data. Here, Max Planck scientists and their collaborators could freely publish their data.
A total of 84 datasets were published through this service in 2020.\footnote{\url{https://s.gwdg.de/G14Bgw}.}\newline

With \href{https://data.goettingen-research-online.de}{GRO.data}, the GWDG offers a service comparable to the Göttingen eResearch Alliance. Here, open research data can be made available via a repository, too. This service, as well as advice on open research data, was one of the services provided by the GWDG to Max Planck researchers in 2020.\newline

With the \href{https://www.wdc-climate.de}{World Data Climate Center}, the German Climate Computing Centre (DKRZ) offers a community-specific data repository. Research data can also be published there as Open Research Data. The Max Planck Institute for Chemistry, for example, made use of this service in 2020 quite far.\footnote{See also \cite[p. 259-260]{wittenburg_fair_2019}.}

\subsection{Max Planck Lighthouse Projects for Open Research Data}
In 2020, the Max Planck Society had a number of Open Research Data projects that have developed quite successfully.\newline

An illustrative example is the FACES\footnote {See \url{https://faces.mpdl.mpg.de}. For more information, see the \cite{ebner_faces_2010}.} platform of the Max Planck Institute for Human Development. The website, developed in 2009 in collaboration with the Max Planck Digital Library (MPDL), offers a collection of high quality images of human faces, grouped by age group, gender and a set of six different facial expressions (emotions).\\
Strictly speaking, this data is not entirely freely available: use of most of the data is restricted to scientific purposes and requires registration/login. However, this unique set of data has proven to be extremely fruitful, with a steady stream of diverse scientific publications over more than a decade.\footnote {Scopus metrics show a steadily growing annual citation rate for FACES since 2010 until today, with a total of more than 700 citations.} \newline

Another very successful project that has been running for a number of years is Movebank\footnote{\url{https://www.movebank.org}.}, which is run by the Max Planck Institute for Animal Behaviour together with various partners around the world. This platform collects and processes current and historical animal movement data from various sources around the world. As most of the data is freely available, it can be used in a variety of ways.\\
With a project start in 2007, Movebank has been around for a long time - with continuous growth and development of the platform. The number of annual publications that can be traced back to Movebank data has also remained high: many hundreds of publications in total.\footnote {\url{https://www.movebank.org/cms/movebank-content/literature}.} \newline

Max Planck's Fritz Haber Institute is heavily involved in the NOMAD Repository \& Archive\footnote{\url{https://nomad-lab.eu/services/repo-arch}.}. Maintained by the Novel Materials Discovery (NOMAD) Laboratory, it is the world's largest repository of input and output files from all major computational materials science programs. The repository contains data in raw format, while the archive provides normalised data in a common, machine-processable format.\\
After starting with the repository in 2014, the project has expanded and developed significantly in the following years as part of the Horizon 2020 European Center of Excellence, NOMAD CoE.\footnote{\cite[p. 260-261]{wittenburg_fair_2019}.}\pagebreak

\section{Publications by the Max Planck Society in the Year 2020}
The previous two chapters show that research data and open access were already widely discussed and in some places already implemented within the Max Planck Society in 2020. At this point in time, there were no general obligations to make research data publicly available. Therefore, this year can be used as a baseline from where to observe future developments. In order to put this understanding on a statistically sound basis, we present in detail our sample of publications by Max Planck scientists in 2020. The focus of this analysis is the handling of research data. It will demonstrate how and where the Max Planck scientists published their data in 2020.  

\subsection{Method}
We manually analysed publications with at least one author from a Max Planck Institute. First, we assessed whether the publication was either empirical\footnote{We defined an empirical paper as one that draws conclusions based on information from the real world. After a long discussion we decided not to consider literature as such information. For example, literature reviews are treated as non-empirical works} or non-empirical, i.e. theoretical. Theoretical publications, which are often found in legal research, but also in mathematics or engineering, are not usually expected to contain research data. We then looked for a statement on data availability or similar. If data was stated to be available, we tried to access the data and categorised the result (all data, some data, no data) and the type of data obtained (raw data, analysed data, summarised data).\footnote{i.e. only tables for figures or single values} If the data availability statement said that the data were "available on request", we contacted the corresponding author (not necessarily from a Max Planck Institute) and asked for the data. We also checked whether the use of software was mentioned and whether this software was available.

\subsection{Approach}
The publication repository of the Max Planck Society \href{https://pure.mpg.de}{MPG.PuRe} contained 15,850 publication references from 2020 (the population, measured on 29/11/2022) that had a publication type for which research data could be expected (journal article, monograph, book, book chapter, edited volume, contribution to edited volume, conference paper, dissertation). Since December 2021, we took a pseudo-randomised\footnote{we created the MD5 hash of the publication identifier and sorted the list by this hash. Later updates were merged into the list} sample of 1,040 publications. We then tried to access the publications, preferably on the publisher's platform. Where this was not possible, we tried to obtain a post-print version from the repository, the author's institute, other libraries or through interlending (n=985). Finally, we looked for a preprint (n=12) or manuscript (n=20). There were 21 publications that we were not able to access despite thorough searches. We excluded these from the sample, which left us with a total N\textsubscript{total}=1,019.

\subsection{Descriptive Analysis}
The vast majority of publications were journal articles (85,3\%), other publication types followed far behind (see table \ref{table:1}).

\begin{table}[H]
\small
\begin{center}
\caption{Publication Types.}
\label{table:1}
\begin{tabular}{ |l|r|r|r| }
\hline
 & Frequency & Percent & in population \\
\hline
Journal article & 869 & 85.3 & 83.3 \\
Thesis & 52 & 5.1 & 5.2 \\
Book chapter & 35 & 3.4 & 4.1\\
Conference paper & 31 & 3.0 & 4.2 \\
Book & 13 & 1.3 & 0.9 \\
Contribution to collected edition & 13 & 1.3 & 1.5 \\
Monograph & 5 & 0.5 & 0.6 \\
Collected edition & 1 & 0.1 & 0.3 \\
\hline
Total & 1019 & 100.0 & 100.0 \\
\hline
\end{tabular}
\end{center}
\end{table}

The Max Planck Society is divided into three sections: the Chemical, Physical and Technical Section (CPTS), the Biological and Medical Section (BMS), and the Humanities and Social Sciences Section (GSHS). Publications are distributed among the sections as shown in the table \ref{table:2}. Combinations of sections reflect collaborations between institutes from different sections.

\begin{table}[h!]
\small
\centering
\caption{MPG sections}
\label{table:2}
\begin{tabular}{ |l|r|r|r| }
\hline
 & Frequency & Percent & in population \\
\hline
CPTS & 555 & 54.7 & 55.6 \\
GSHS & 245 & 23.9 & 23.4 \\
BMS & 214 & 20.0 & 19.4 \\
CPTS,BMS & 2 & 1.1 & 1.4 \\
Central & 2 & 0.2 & 0.1 \\
GSHS,BMS & 1 & 0.1 & 0.0 \\
\hline
Total & 1019 & 100.0 & 100.0 \\
\hline
\end{tabular}
\end{table}

\begin{figure}[H]
    \centering
    \resizebox{0.8\textwidth}{!}{%
\begin{tikzpicture}
\begin{axis}[ybar,
            width=\textwidth,
            bar width=6mm,
            symbolic x coords={Springer Nature, Wiley, Elsevier, American Chemical Society, American Physical Society, IOP, Oxford University Press, Springer, Royal Society of Chemistry, EDP Sciences},
            xtick=data,
            ylabel=Percent of Total Selected Publications,
            xticklabel style={rotate=45,anchor=north east},
            ymajorgrids,
            nodes near coords,
            ymin=0,
            axis y line*=left,
            axis x line*=bottom
          ]
    \addplot [fill={verde}] coordinates {
        (Springer Nature, 7.2) 
        (Wiley, 7.2) 
        (Elsevier, 5.8) 
        (American Chemical Society, 4.7)
        (American Physical Society, 4.5)
        (IOP, 4.2)
        (Oxford University Press, 3.8)
        (Springer, 3.1)
        (Royal Society of Chemistry, 2.5)
        (EDP Sciences, 2.3)
            };
\end{axis}
\end{tikzpicture}
}
    \caption{Ten Most Frequent Publisher of the Empirical Selected Publications}
    \label{fig:publisher-1}
\end{figure}
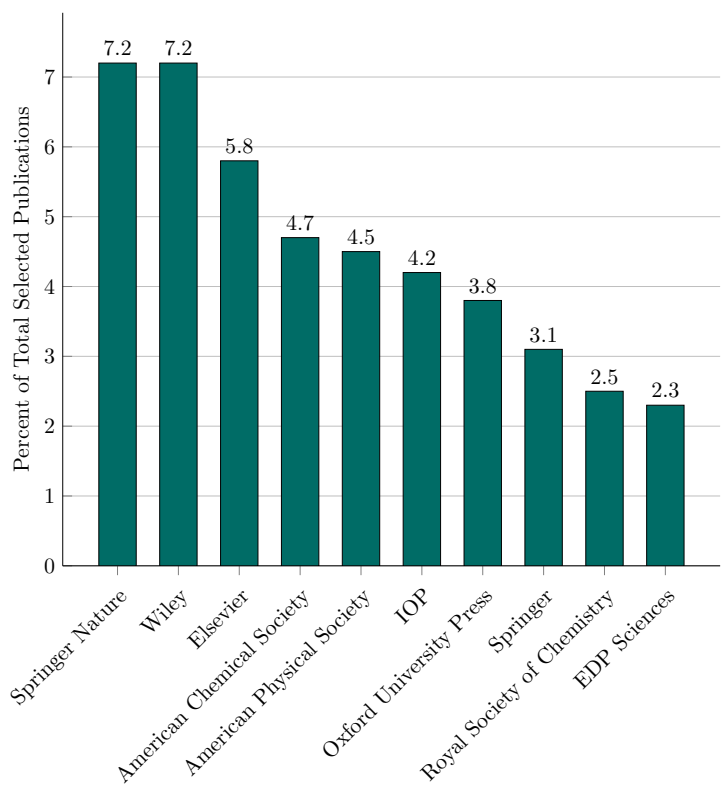

In 2020, Max Planck scientists published their research results in many different places. The absolute numbers of all publications by Max Planck scientists certainly show similarities to the national figures provided by the \href{https://esac-initiative.org/market-watch/}{ESAC Market Watch} for Germany.\footnote{For more details and the code see mainly \cite{jahn_subugoeoa2020cadata_2019}.} For the publications selected in this study, this means that the top three publishers were Springer Nature, Elsevier and Wiley. As figure \ref{fig:publisher-1} shows, in 2020, one fifth of the selected empirical publications were published at the top three.\newline

\subsection{Selected Publications}
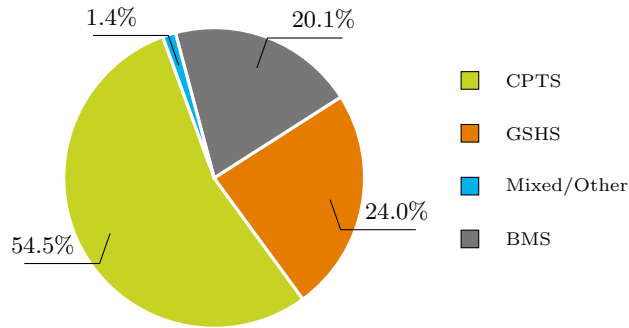
\begin{figure}[H]
\centering
\begin{tikzpicture}
[
    pie chart,
    slice type={comet}{blu},
    slice type={legno}{rosso},
    slice type={coltello}{giallo},
    slice type={sedia}{viola},
    pie values/.style={font={\small}},
    scale=2.0
]

\pie{}{54.5/coltello,24.0/legno,20.1/sedia,1.4/comet}


\legend[shift={(1.7cm,1cm)}]{{CPTS}/coltello,{GSHS}/legno, {Mixed/Other}/comet,{BMS}/sedia}
\end{tikzpicture}
\caption{Distribution of Publications by Max Planck Sections}
\label{fig:sections}
\end{figure}

We analysed a total of 1,019 data publications in more detail. The CPTS accounted for slightly more than half of this sample, see figure \ref{fig:sections}. This result is to be expected due to the relatively large number of institutes and staff in this section. In addition, about a quarter and a fifth of the publications came from the BMS and GSHS, respectively. Collaborations involving Max Planck scientists from several sections account for only 1.4\% of the total number of publications.

\begin{figure}[h!]
\centering
\begin{tikzpicture}
[
    pie chart,
    slice type={comet}{blu},
    slice type={legno}{rosso},
    slice type={coltello}{giallo},
    slice type={sedia}{viola},
    slice type={caffe}{verde},
    pie values/.style={font={\small}},
    scale=2.0
]


\pie{}{57.38/caffe,14.34/legno,18.44/sedia,9.84/coltello}

\legend[shift={(1.7cm,1cm)}]{{Humanities}/legno,{Human Sciences}/caffe, {Social Sciences}/coltello,{Legal Sciences}/sedia}
\end{tikzpicture}
\caption{{Distribution of Publications within the GSH Section}}
\label{fig:GSH-distribution}
\end{figure}
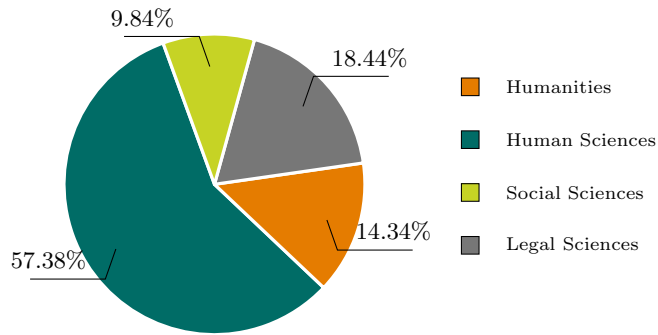

An explicit look at the GSH section with its 244 publications is particularly remarkable in this distribution. Here, comparatively different disciplinary cultures are linked. The more detailed distribution in figure \ref{fig:GSH-distribution} shows the distribution of research fields. The majority of GSHS publications in 2020 came from the human sciences. This is followed at a distance by publications from the fields of law and the humanities.\footnote{Our internal breakdown of the GSH section can be found in the corresponding data publication, see \ref{data-and-code-from-us}.} Publications in the social sciences have the smallest share in the sample.

The core of a bibliometric analysis of research data is mentioned in textual publications. However, it is important to remember that theoretical work does not necessarily have to include or be based on research data. Behaviours and characteristics can be postulated on the basis of theoretical assumptions and deductions. These would only have to be supported or refuted by empirical data, for example, in a second step. For this reason, to gain this insight, we inquired to what extent each publication could be considered theoretical. This classification is important in the following, as we only analyse the handling of research data in 2020 in the case of non-theoretical publications. This is because we limit ourselves to dealing with the management of research data. The inclusion of theoretical works, which are not supposed to have research data, would put the results of the analysis into an additional negative context.\newline

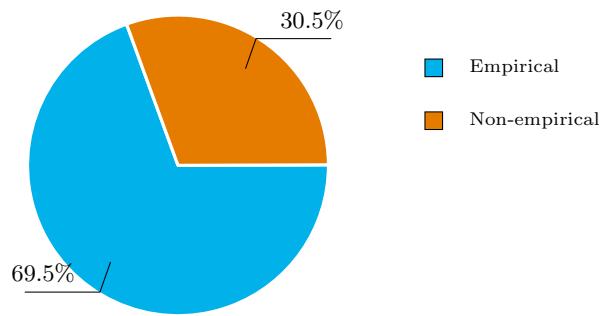
\begin{figure}[h!]
\centering
\begin{tikzpicture}
[
    pie chart,
    slice type={comet}{blu},
    slice type={legno}{rosso},
    slice type={coltello}{giallo},
    slice type={sedia}{viola},
    pie values/.style={font={\small}},
    scale=2.0
]

\pie{}{69.5/comet,30.5/legno}
\legend[shift={(1.7cm,1cm)}]{{Empirical}/comet,{Non-empirical}/legno}
\end{tikzpicture}
\caption{Classification According to Empirical and Non-Empirical}
\label{fig:empirical}
\end{figure}

According to this classification shown in figure \ref{fig:empirical}, about 30\% of the randomly selected papers are non-empirical. In contrast N\textsubscript{empirical} = 708 are empirical publications.

\begin{figure}[h!]
    \centering
    \resizebox{0.8\textwidth}{!}{
        \begin{tikzpicture}
            \begin{axis}
                [ybar stacked,
                        width=\textwidth,
                        bar width=6mm,
                        symbolic x coords={BMS,CPTS,Human Sciences,Social Sciences,Legal Sciences,Humanities,Mixed/Other},
                        xtick=data,
                        ylabel=Percent of Total Publications,
                        xticklabel style={rotate=45,anchor=north east},
                        ymajorgrids,
                        nodes near coords,
                        ymin=0,
                        axis y line*=left,
                        axis x line*=bottom
                ]
                \addplot+[ybar] plot [fill={adjusted_red}, text=white] coordinates
                {
                    (BMS, 37)
                    (CPTS, 137)
                    (Human Sciences, 44)
                    (Social Sciences, 16)
                    (Legal Sciences, 38)
                    (Humanities, 35)
                    (Mixed/Other, 4)
                };
                \addplot+[ybar] plot [fill={verde}, text=white] coordinates
                {
                    (BMS, 168)
                    (CPTS, 418)
                    (Human Sciences, 96)
                    (Social Sciences, 8)
                    (Legal Sciences, 7)
                    (Humanities, 0)
                    (Mixed/Other, 11)
                };
                \legend{Theoretical,Empirical}
            \end{axis}
        \end{tikzpicture}
    }
    \caption{Empirical and theoretical publications in different domains}
    \label{fig:empirical-domains}
\end{figure}
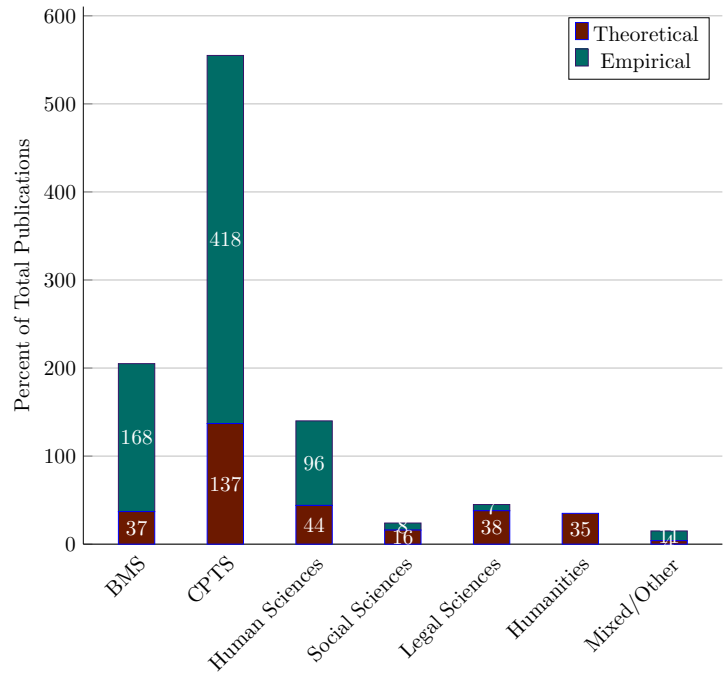

In relation to the scientific domains described above, the highest ratio of empirical publications are in biological/medical research (82\%) whereas in humanities, no empirical works appear in this sample (see figure \ref{fig:empirical-domains}). We excluded theoretical work from the sample.

\pagebreak

\subsection{Results}

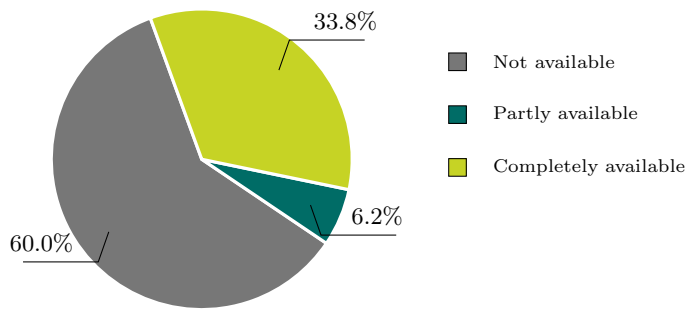
\begin{figure}[h!]
\centering
\begin{tikzpicture}
[
    pie chart,
    slice type={comet}{blu},
    slice type={legno}{rosso},
    slice type={coltello}{giallo},
    slice type={sedia}{viola},
    slice type={caffe}{verde},
    pie values/.style={font={\small}},
    scale=2.0
]

\pie{}{60.0/sedia,6.2/caffe,33.8/coltello}

\legend[shift={(1.7cm,1cm)}]{{Not available}/sedia,{Partly available}/caffe, {Completely available}/coltello}
\end{tikzpicture}
\caption{Research Data Availability within Selected Empirical Publications}
\label{fig:selected-empirical}
\end{figure}

Figure \ref{fig:selected-empirical} shows our sample, where 40\% of the empirical publications provide research data. Conversely, this means that research data is not available in 60\% of the publications with empirical focus.

\begin{figure}[ht]
    \centering
    \resizebox{0.7\textwidth}{!}{%
        \begin{tikzpicture}
            \begin{axis}[ybar,
                        width=\textwidth,
                        bar width=6mm,
                        symbolic x coords={No, Yes, After request, Partly, After login, After payment, Other},
                        xtick=data,
                        ylabel=Numbers of Publications,
                        xticklabel style={rotate=45,anchor=north east},
                        ymajorgrids,
                        nodes near coords,
                        ymin=0,
                        axis y line*=left,
                        axis x line*=bottom
                      ]
                \addplot [fill={verde}] coordinates {
                    (No, 425) 
                    (Yes, 137) 
                    (After request, 84) 
                    (Partly, 43)
                    (After payment, 8)
                    (After login, 10)
                    (Other, 1)
                        };
            \end{axis}
        \end{tikzpicture}
    }
    \caption{Types of Availability.}
    \label{fig:types_avail}
\end{figure}
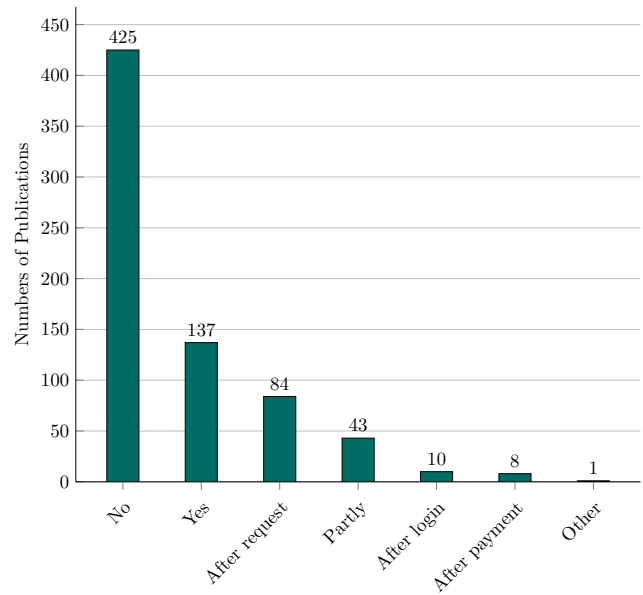

However, the claimed 40\% of available research data N\textsubscript{empirical} is even more limited. As figure \ref{fig:selected-empirical} visualises, open research data is actually completely available for only 137 publications. This is partially the case for a further 43 (6,2\%) publications. For a total of 84 publications it is stated that the research data will be made available upon "reasonable request". In theory, these publications should be accessible, but in reality this is not always the case. This phenomenon will be discussed in more detail in the following chapter \ref{Available-upon-request}. 
To anticipate this, in our case only a fraction of the research data was actually made available; as shown in figure  \ref{fig:types_avail}.

The handling of research data in theses, especially doctoral theses, is striking: Only 18.2\% of the 44 theses have (some) research data, whereas 81.8\% have none.\newline

\begin{table}[H]
\small
\begin{center}
\caption{Available data per publication type with $n\ge5$}
\label{table:4}
\begin{tabular}{ |l|r|r|r|r| }
\hline
 & partly(\%) & complete(\%) & sum(\%) & n \\
\hline
Journal article & 6,2 & 36,0 & 42,3 & 641 \\
Conference paper & 0,0 & 23,5 & 23,5 & 17 \\
Thesis & 6,8 & 11,4 & 18,2 & 44 \\
\hline
Total & 6,1 & 33,9 & 40,0 & 702 \\
\hline
\end{tabular}
\end{center}
\end{table}

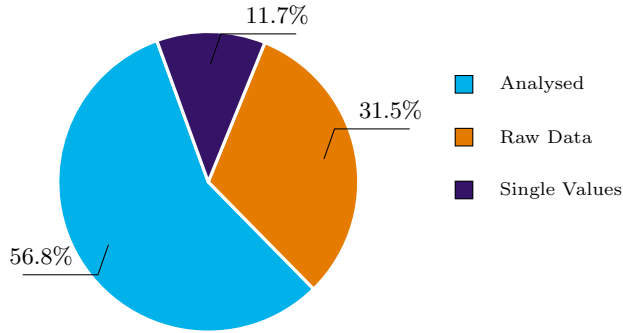
\begin{figure}[h!]
    \centering
    \begin{tikzpicture}
        [
            pie chart,
            slice type={comet}{blu},
            slice type={legno}{rosso},
            slice type={coltello}{giallo},
            slice type={sedia}{viola},
            slice type={reddy}{red},
            pie values/.style={font={\small}},
            scale=2.0
        ]
        \pie{}{56.8/comet,31.5/legno,11.7/reddy}
        \legend[shift={(1.7cm,1cm)}]{{Analysed}/comet,{Raw Data}/legno, {Single Values}/reddy}
    \end{tikzpicture}
    \caption{Data aggregation level}
    \label{fig:dataaggregationl}
\end{figure}

At the same time, a closer look at the aggregation level of the data in empirical publications shows that   31.5\% of the research data is in its raw state, see figure \ref{fig:dataaggregationl}. In contrast, only 11.7\% of research data consists of a series of individual values. Most of the publications (56.8\%) offer research data in an analysed form.\newline

The availability of data also varies widely among the publishers that released the works, shown in table \ref{table:3}. Data is mostly available in high-impact journals like Cell and Nature and in the open-access journals of Frontiers and Copernicus. At the lower end of the scale are the learned societies with the exception of the American Geophysical Union where data is available for every publication. However, data policies can be an important point for clarifying these differences. Publishers have very different policies on data availability and reproducibility. As shown by Canadian colleagues, data policies have become increasingly common in journals since the mid-2010s.\footnote{\cite[pp. 381-382]{dearborn_changing_2018}.} It seems quite plausible that there is a causal relationship between data availability and the existence of a publisher's data policy.

\begin{table}[H]
\small
\begin{center}
\caption{Available data per publisher with $n \ge 5$}
\label{table:3}
\begin{tabular}{ |l|r|r|r|r| }
\hline
 & partly(\%) & complete(\%) & sum(\%) & n \\
\hline
American Geophysical Union & 0.0 & 100.0 & 100.0 & 8 \\
Cell Press & 20.0 & 73.3 & 93.3 & 15 \\
Frontiers & 0.0 & 90.9 & 90.9 & 11 \\
Copernicus Publications & 16.7 & 66.7 & 83.3 & 12 \\
Nature & 9.8 & 68.6 & 78.4 & 51 \\
Academic Press & 25.0 & 50.0 & 75.0 & 8 \\
BioMed Central & 0.0 & 71.4 & 71.4 & 7 \\
Oxford University Press & 7.4 & 51.9 & 59.3 & 27 \\
AAAS & 0.0 & 50.0 & 50.0 & 10 \\
American Institute of Physics & 20.0 & 20.0 & 40.0 & 5 \\
MDPI & 20.0 & 20.0 & 40.0 & 10 \\
Wiley & 7.8 & 29.4 & 37.3 & 51 \\
Springer & 4.5 & 31.8 & 36.4 & 22 \\
Cambridge University Press & 0.0 & 33.3 & 33.3 & 6 \\
EDP Sciences & 6.3 & 25.0 & 31.3 & 16 \\
Pergamon & 0.0 & 28.6 & 28.6 & 7 \\
Royal Society of Chemistry & 0.0 & 22.2 & 22.2 & 18 \\
Elsevier & 0.0 & 22.0 & 22.0 & 41 \\
American Chemical Society & 9.1 & 6.1 & 15.2 & 33 \\
IOP & 3.3 & 10.0 & 13.3 & 30 \\
American Physical Society & 0.0 & 3.1 & 3.1 & 32 \\
\hline
Total & 6.4 & 36.2 & 42.6 & 420 \\
\hline
\end{tabular}
\end{center}
\end{table}

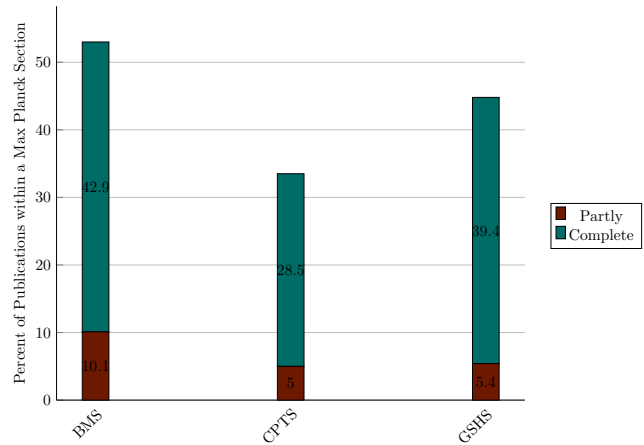
\begin{figure}[h!]
    \centering
    \resizebox{0.7\textwidth}{!}{%
\begin{tikzpicture}
\begin{axis}[ybar stacked,
            width=\textwidth,
            bar width=6mm,
            symbolic x coords={BMS, CPTS, GSHS, BMS and CPTS},
            xtick=data,
            ylabel=Percent of Publications within a Max Planck Section,
            xticklabel style={rotate=45,anchor=north east},
            ymajorgrids,
            ymin=0,
            nodes near coords,
            axis y line*=left,
            axis x line*=bottom,
            legend style={at={(1.25,0.5)},= east}
          ]
    \addplot[fill={adjusted_red}] coordinates {
        (BMS, 10.1)
        (CPTS, 5.0) 
        (GSHS, 5.4)
        };
    \addplot [fill={verde}] coordinates {
        (BMS, 42.9)
        (CPTS, 28.5) 
        (GSHS, 39.4) 
        };
    \legend{Partly, Complete}
\end{axis}
\end{tikzpicture}
}
    \caption{Data Availability within the Max Planck Sections of the Empirical Selected Publications}
    \label{fig:publications-MPG-sec}
\end{figure}

Similar to the different publishers, there are differences in data availability between the sections of the Max Planck Society. In terms of data availability within the three sections, an average of 33.8\% (N\textsubscript{Empirical with data}=239) of empirical publications are linked to accessible data. An average of 6.2\% (N\textsubscript{Empirical with partly data}=44) of the publications in the sections have partial data. The figure \ref{fig:publications-MPG-sec} shows the statistical variation between the sections. The data availability within the CPTS is slightly below average, while the BMS and the GSHS are slightly above average. Combining this with the values for partial data availability, we see that within the BM section almost every second publication contains research data. In the CPT section, however, this is the case for every third publication.\newline

At the same time, it is important to have a look at the similarities in these figures. The difference between the BMS and the CPTS is only 15\%. As a result it can be argued that data availability within the Max Planck Society differs only slightly between the individual sections. No section outperforms another by a multiple in terms of percentage data availability. A long-term perspective is particularly interesting for these figures. Here it might be possible to observe whether different developments and internal discussions regarding the availability of research data in publications take place in different subject groups or times.

\subsection{Data Licenses}

The FAIR principles mentioned in chapter \ref{After-2016} have increasingly become a quasi-standard. The key idea behind these principles is to make the descriptive aspects of data explicit. This concerns, for example, a statement about the possibility of using the data, which is made explicit by convention through a licence. It is therefore an interesting aspect to ask about the use of licences within the here presented Max Planck sample. This provides a better perspective on the concrete application of the FAIR principles, independent of the often communicated relevance of the principles.\newline

If we ask this question of the 239 available publications whose research data we were able to access, the result is somewhat surprising. Three quarters of the data do not have their own licence. It is therefore unclear how and under what conditions the data could be reused. In contrast, only a quarter of the data had an explicit licence. Of these, the CC BY 4.0 licence was clearly the most widely used, at 15\%. CC0 is another licence worth mentioning, with two percent. The remaining, many licences are lost in a kind of background noise at 6.8\%. 

\subsection{Data Repositories}

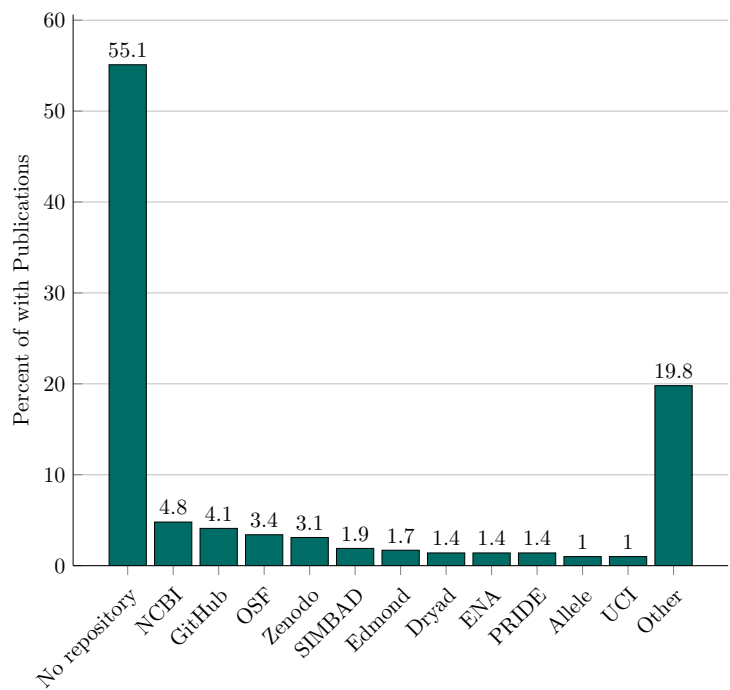
\begin{figure}[h]
    \centering
    \resizebox{0.8\textwidth}{!}{%
\begin{tikzpicture}
\begin{axis}[ybar,
            width=\textwidth,
            bar width=6mm,
            symbolic x coords={No repository, NCBI, GitHub, OSF, Zenodo, SIMBAD, Edmond, Dryad, ENA, PRIDE, Allele, UCI, Other},
            xtick=data,
            ylabel=Percent of with Publications,
            xticklabel style={rotate=45,anchor=north east},
            ymajorgrids,
            nodes near coords,
            ymin=0,
            axis y line*=left,
            axis x line*=bottom
          ]
    \addplot  [fill={verde}] coordinates {
        (No repository, 55.1)
        (NCBI, 4.8) 
        (GitHub, 4.1) 
        (OSF, 3.4)
        (Zenodo, 3.1)
        (Edmond, 1.7)
        (Dryad, 1.4)
        (ENA, 1.4)
        (PRIDE, 1.4)
        (Allele, 1.0)
        (SIMBAD, 1.9)
        (UCI, 1.0)
        (Other, 19.8)
        };
\end{axis}
\end{tikzpicture}
}
    \caption{Data Repositories at the Available Research Data}
    \label{fig:repos}
\end{figure}

As with licences, looking at the used data repositories in our sample reveals a behaviour in relation to data availability. Here, it is particularly interesting to see which category of repository is represented. Figure \ref{fig:repos} shows the distribution of publications by repository: the first value is striking. However, it should be noted that the use of no repository rather indicates that the data was for example attached directly to the publication. Data papers as a genre are also possible. So there are other ways of making data available than using data repositories.\newline

The largest number of Max Planck researchers (4.8\%) of the publications sample have made their data available via the National Center for Biotechnology Information. This US data repository specialises in the storage of molecular biology data. It has become the quasi-standard repository for data in this field of research. Despite its institutional reference, it can be classified as a subject specific repository due to its focus on specific topics. In addition, ENA, PRIDE and Allele have been used as other subject-specific repositories for the availability of research data. They all have a focus on bioinformatics. In this research environment it can be observed that a research culture for the use of repositories already exists. Some of the repositories are even internally differentiated, such as proteomics data in PRIDE. In addition, SIMBAD stands out as a subject-specific repository for astronomical research data. After all, almost 2.0\% of all research data in our example was published there.\newline

In addition to subject-specific repositories, the use of generic data repositories by Max Planck scientists can be observed in 2020. Figure \ref{fig:repos} shows that they were used frequently in our sample. It is remarkable that \url{https://github.com}, a primary software repository, was used as a data repository in 4.1\% of cases. The institutional data repository of the Max Planck Society, Edmond, had been used (1.7\%).

\pagebreak

\subsection{Data ``Available'' Upon Request} \label{Available-upon-request}

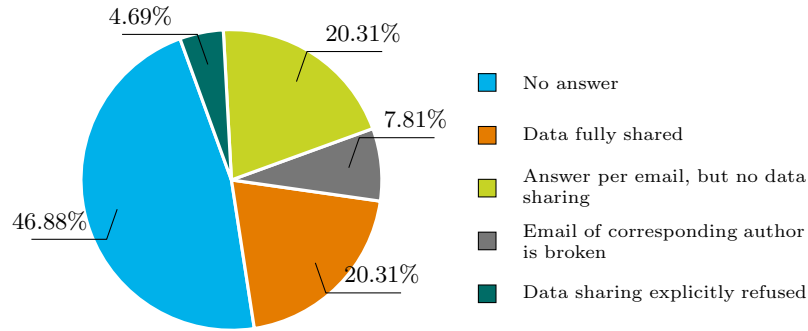
\begin{figure}[h!]
\centering
\begin{tikzpicture}
[
    pie chart,
    slice type={comet}{blu},
    slice type={legno}{rosso},
    slice type={coltello}{giallo},
    slice type={sedia}{viola},
    slice type={caffe}{verde},
    pie values/.style={font={\small}},
    scale=2.0
]

\pie{}{46.88/comet,20.31/legno,7.81/sedia,20.31/coltello,4.69/caffe}

\legend[shift={(1.7cm,1cm)}]{{No answer}/comet,{Data fully shared}/legno, {Answer per email, but no data sharing}/coltello,{Email of corresponding author is broken}/sedia, {Data sharing explicitly  refused}/caffe}
\end{tikzpicture}
\caption{Distribution of responses to a data request}
\label{fig:available-not-sharing}
\end{figure}

Data is not always readily available. As discussed in chapter \ref{available-statements-text}, this phenomenon is associated with some problems. In our sample, this is exemplified by the 84 publications for which data are described as available upon (reasonable) request. It was possible to obtain data on request from about 20\% of the publications. Specifically, about 8\% of the corresponding authors could not be contacted because their 2020 email address no longer worked and about half of the authors did not answer at all. At the same time it was also observed that, despite justified requests, the data was not available. Figure \ref{fig:available-not-sharing} visualises this.
    
In fact, it is alarming that more than three-quarters of the research data claimed to be available are in fact unavailable. With 84 publications related to this, the data base is comparatively small. Nevertheless, it can be said that the enquiries and correspondence leading up to the final result were lengthy, time-consuming and often frustrating. Similar to other studies described above, our experience shows that obtaining data usually involves a long email correspondence with the authors.\footnote{\cite[p. 8]{tedersoo_data_2021}.}

\subsection{Research Software in Publications}

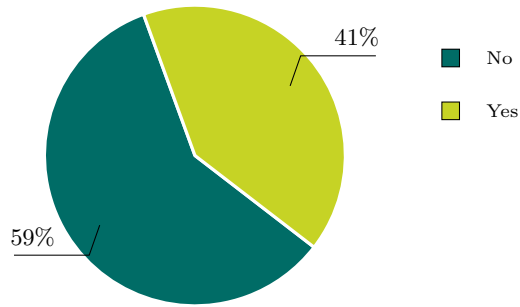
\begin{figure}[h!]
\centering
\begin{tikzpicture}
[
    pie chart,
    slice type={comet}{blu},
    slice type={legno}{rosso},
    slice type={coltello}{giallo},
    slice type={sedia}{viola},
    slice type={caffe}{verde},
    pie values/.style={font={\small}},
    scale=2.0
]

\pie{}{59/caffe,41/coltello}
\legend[shift={(1.7cm,1cm)}]{{No}/caffe,{Yes}/coltello}
\end{tikzpicture}
\caption{Mentioned Software in the Selected Publications}
\label{fig:software}
\end{figure}

Software is often needed to reproduce the analysis of research data and to be able to follow the path of scientific knowledge. Research software was mentioned in some form in just over 41\% of the selected publications (see figure \ref{fig:software}). Research software was only marginally the focus of this study. Nevertheless, it is remarkable how clear the treatment of the accessibility in comparison to research data is.

Of particular note is the more explicit use of research software in publications compared to research data. Research software is mentioned in 288 of the N\textsubscript{total} = 708 selected empirical publications. And of these 288 software packages, 226 -- i.e. 78\% -- are openly accessible. This is a significant difference compared to the figures from openly available research data.\footnote{However, it should not be concealed that even with research software not all elements are always accessible and sharing with others is refused here. Overall, however, this is on a much lower level. See also the observations from \cite[832]{assel_statistical_2018}} This is in line with other research on the open availability of research software. For example, in 2021, research software on \url{https://github.com} was already mentioned in over 20\% of all publications on \url{https://arxiv.org/}.\footnote{\cite{escamilla_rise_2022}.}

\section{Discussion}
"\textit{There is nothing either good or bad, but thinking makes it so.}"\footnote{William Shakespeare (1564–1616), Hamlet, act II, scene 2.} To cite Shakespeare, how we value open research data depends on our own point of view and context. The selection of publications from 2020 presented here can therefore also be discussed from different perspectives on the part of the Max Planck Society. We have identified five key aspects from the analysis of publications in 2020 and its relation to data availability.\newline

1. Expectations and actual results differ significantly by \textbf{type of data availability}. Particularly in empirical work one would expect the research data to be available in some way, whether as open research data or with restricted access. Since in terms of good scientific practice, it should be possible to reproduce the results. Considering this statement, only 40\% of research data available or partially available for empirical publications is not much. Theoretically, one could expect 100\% availability. In all previous studies, however, the found numbers were at a very low level.\footnote{See chapter \ref{After-2016}.} Open or restricted access -- for which there are good reasons -- would be irrelevant for the time being. Reproducibility would mainly be guaranteed in both cases. However, almost 60\% of empirical work is without data. For an excellent research organisation like the Max Planck Society, which sees itself as one of the leading scientific organisations operating in the field of basic research, there is still potential for a greater data availability.\newline

2. The \textbf{aggregation} of the available data is predominantly analysed. But there is also raw data. There is a discussion about laboratory data and how far it can or should be published.\footnote{See for example \cite[p. 192]{pinel_caring_2020}.} There is therefore no definite answer to the question of when data should be published.  This is probably a case-by-case decision. For the Max Planck Society in its diversity, it does not seem sensible to formulate a general rule for the degree of aggregation.
\newline

3. A \textbf{data policy} can increase the availability of data associated with textual publications. Such normative requirements on the part of publishers lead and will increasingly lead to available data in the near future. It can be assumed that publishers, funders and scientific organisations will increasingly develop such normative frameworks for scientists. However, this is not the same as demanding open access to research data. There may be good reasons for publishing data with restricted access, for example in clinical trials or industrial contract work.\\
"\textit{Nothing can come of nothing.}"\footnote{William Shakespeare (1564–1616), King Lear, act I, scene 1.} Greater visibility of their research results and the application of Open Science can, for example, be motivation for a data policy. Such a normative framework is indispensable if the Max Planck Society or individual institutes want to motivate their own scientists in this direction. Recent developments at the Helmholtz Association in particular show how such a path could be taken in German basic research.\footnote{See for example \cite{helmholtz_open_science_office_helmholtz_2022} and \cite{ferguson_indikatoren_2021}.}
\newline

4. The concept of \textbf{data availability statements} like "Data available on reasonable request" do not work as expected. Response rates are low. The communication effort is usually high. Both our experience and other studies have shown that there is a mismatch between effort and return. Storing the research data in a data repository -- with open or restricted access -- would eliminate this problem. \\
For the Max Planck Society, this may lead to a kind of recommendation that research data should be published in suitable infrastructures. These could be research data repositories, data journals or similar solutions.
\newline

5. The data handling culture of the \textbf{sections of the Max Planck Society} do differ from each other. At the same time, the differences in the publication of research data are not as great as one might expect. This suggests that the Max Planck Society has already taken aspirations towards a data sharing culture. For example, the new rules on good scientific practice mention research data and the explicit handling of them quite often.\footnote{A brief overview of the research data aspect of the new rules can be found in \cite{grossmann_gwp_2021}.} However, if we compare the Max Planck Society with, for example, the Helmholtz Association and its use of research software, or the Charité and its open research applications, there is still a lot of potential that can be employed within the Max Planck Society.\footnote{\cite{helmholtz-gemeinschaft_empfehlungen_2019}, \cite{iarkaeva_semi-automated_2022} and \cite{vladislav_nachev_charite_2023}.}\pagebreak

\section{Perspectives}

Analysing the publications by Max Planck researchers in 2020 is the past. What has happened since then?\newline

Open Science and Open Research Data are becoming increasingly relevant in the German research landscape. This is clearly indicated, for example, by the positioning of the DFG in autumn 2022.\footnote{\cite[pp. 4-5]{forschungsgemeinschaft_open_2022}.} There are many advantages to sharing data and code for a culture of science.\footnote{See also the detailed discussion of the advantages and disadvantages in \cite{gomes_why_2022}.} With a focus on Germany, the NFDI will play an increasing role in this. One of its first successes is that research data and its management have not only reached the scientific community but German decision-makers, e.g. in politics. It is therefore not a far-fetched thesis to assume that research data will become an increasingly important topic in the coming years. The debate has already lost some of its drama. It is now less a question of "why" and more a question of "how".\newline

All this manifests itself in the fact that the topic of research data is increasingly being dealt with locally at the Max Planck Institutes. The new guidelines on good scientific practice from the end of 2022 also made a significant contribution to this.\footnote{\cite{max-planck-gesellschaft_verhaltensregeln_2022}.} There are local working groups, initiatives and committed colleagues looking for local solutions in their departments and institutes. At the same time, a Max Planck-wide network of RDM experts is evolving. Events such as the regular RDM workshops of the Max Planck Digital Library are evidence of this.\footnote{See \url{https://rdm.mpdl.mpg.de/mpdl-services/workshops/}.} However, it remains to be seen whether the developments at the individual institutes will be merged. It also remains to be seen whether concepts such as "data stewardship" will become established in individual disciplines.\newline

There is already a clear need for more knowledge about research data within one's own institution. Services such as the \href{https://quest-dashboard.charite.de}{Charité Dashboard on Responsible Research} mentioned above, the \href{https://fairdashboard.helmholtz-metadaten.de/}{HMC Dashboard on Open and FAIR Data in Helmholtz} or the \href{https://barometredelascienceouverte.esr.gouv.fr}{French Open Science Monitor} provide a first glimpse of what such bibliometric services might look like. The transition to evaluation methods and research assessment is, of course, imminent. Since the Max Planck Society has a real interest in such metrics, it would certainly be in everyone's interest to be able to offer such services (internally) in the long term. Such dashboards with a focus on research data can, for example, also open up longitudinal perspectives on how Max Planck scientists deal with research data. Nevertheless, it is clear that such analyses of data availability as we have presented here should be carried out more frequently. It will be interesting to see how publishing behaviour changes as a result of the developments mentioned.
In the end, however, it is said, "[b]\textit{e great in act, as you have been in thought.}"\footnote{William Shakespeare (1564–1616), King John, act V, scene 1.}\pagebreak

\section{Abbreviations}
\begin{itemize}
\item BMC = BioMed Central
\item BMS = Biological and Medicine Section of the Max Planck Society
\item CPTS = Chemistry, Physics and Technology Section of the Max Planck Society
\item DFG = Deutsche Forschungsgemeinschaft
\item DMP= Data Management Plan
\item DOI = Digital Object Identifier
\item ENA = European Nucleotide Archive
\item EOSC = European Open Science Cloud
\item ESAC = Efficiencies and Standards for Article Charges
\item FAIR = Acronym of findability, accessibility, interoperability, and reusability regarding data principles, see \cite{wilkinson_fair_2016}
\item GSHS = Human Sciences Section of the Max Planck Society
\item GWK = Joint Science Conference
\item HRK = Hochschulrektorenkonferenz (German Rectors' Conference)
\item MPCDF = Max Planck Computing Data Facility
\item MPDL = Max Planck Digital Library
\item MPI = Max Planck Institute
\item MPS = Max Planck Society
\item NCBI = National Center for Biotechnology Information
\item NFDI = National Research Data Infrastructure
\item OECD = Organization for Economic Co-operation and Development
\item ORDP = Open Research Data Pilot
\item CC = Creative Commons
\item PLOS = Public Library of Science
\item PRIDE = Proteomics Identifications Database
\item SIMBAD = Set of Identifications, Measurements and Bibliography for Astronomical Data
\item URL = Uniform Resource Locator
\item WR = German Science and Humanities Council (Wissenschaftsrat)
\end{itemize}

\pagebreak
\section{Statements and Comments}
\subsection{Data and Code Availability Statement} \label{data-and-code-from-us}
The data and code are freely available in Edmond via \url{https://doi.org/10.17617/3.XI0LP5}. Via Edmond the scripts can also run directly in Mybinder using \url{https://mybinder.org/v2/dataverse/10.17617/3.XI0LP5}.

\subsection{Author contributions}
The study design was set up by Franke and Grossmann. The drawing of the sample was programmed by Franke. Ho and Matthiesen were mainly responsible for data collection and research. Ho and Matthiesen were also in charge of the graphical representations of the results. Ho and Franke have evaluated and analysed the data collected. The texts for the open data examples were written by Boosen. Other text elements including the bibliography were written by Grossmann. Corrections and linguistic adjustments were done by Leiminger.

\pagebreak

\printbibliography
\end{document}